\documentclass[12pt,preprint]{aastex}
\usepackage{epsf}

\begin{document}

\def\plotsix#1#2#3#4#5#6{\centering \leavevmode
\epsfxsize=.35\columnwidth \epsfbox{#1} \vspace{.25in}
\epsfxsize=.35\columnwidth \epsfbox{#2} \hfil
\epsfxsize=.35\columnwidth \epsfbox{#3} \vspace{.25in}
\epsfxsize=.35\columnwidth \epsfbox{#4} \hfil
\epsfxsize=.35\columnwidth \epsfbox{#5} \vspace{.25in}
\epsfxsize=.35\columnwidth \epsfbox{#6}}

\def\etal{{\it et al.~}}
\def\eg{{\it e.g.,~}}
\def\ie{{\it i.e.~}}

\title{Numerical Study of Compressible Magnetohydrodynamic Turbulence
in Two Dimensions}

\author{Hyesook Lee\altaffilmark{1},
        Dongsu Ryu\altaffilmark{1},
        Jongsoo Kim\altaffilmark{2},
        T. W. Jones\altaffilmark{3},
    and Dinshaw Balsara\altaffilmark{4}}

\altaffiltext{1}
{Department of Astronomy \& Space Science, Chungnam National University,
Daejeon 305-764, Korea: lhs@cano\-pus.chungnam.ac.kr,
ryu@canopus.chungnam.ac.kr}
\altaffiltext{2}
{Korea Astronomy Observatory, 61-1, Hwaam-Dong, Yusong-Ku, Taejon 305-348,
Korea: jskim@kao.re.kr}
\altaffiltext{3}
{Department of Astronomy, University of Minnesota, Minneapolis, MN 55455:
twj@msi.umn.edu}
\altaffiltext{4}
{Department of Physics, University of Notre Dame, Notre Dame, IN 46556:
dbalsara@nd.edu}

\begin{abstract}

We have studied forced turbulence of compressible magnetohydrodynamic
(MHD) flows through two-dimensional simulations with different
numerical resolutions. First, hydrodynamic turbulence with Mach number
$\langle M_s \rangle_{\rm init} \equiv \langle v
\rangle_{\rm rms}/ c_s = 1$ and density compression
${\langle \delta\rho / \rho \rangle}_{\rm rms} \simeq 0.45$
was generated by enforcing a random force. Then, initial,
uniform magnetic fields of various strengths were added
with Alfv\'enic Mach number $\langle M_A \rangle_{\rm init}
\equiv \langle v \rangle_{\rm rms} / c_{A, {\rm init}} \gg 1$.
An isothermal equation of state was employed, and no explicit dissipation
was included. In our simulations, the maximum amplification
factor of magnetic energy depends on resolution and is proportional
to $n_x^{1.32}$, where $n_x$ is
the number of grid cells spanned by the computational box size.
After the MHD turbulence is saturated, the resulting flows are
categorized as very weak field (VWF), weak field (WF), and strong
field (SF) classes, which have $\langle M_A \rangle \equiv
\langle v \rangle_{\rm rms} / \langle c_A \rangle_{\rm rms} \gg 1$,
$\langle M_A \rangle > 1$, and $\langle M_A \rangle \sim 1$,
respectively. The flow character in the VWF cases is similar to
that of hydrodynamic turbulence. In the WF cases, magnetic energy
is still smaller than kinetic energy in the global sense, but
magnetic field can become locally important. Hence, not only in
the SF regime but also in the WF regime, turbulent transport is
suppressed by the magnetic field. In the SF cases, the energy power
spectra in the inertial range, although no longer power-law,
exhibit a range with slopes close to $\sim1.5$, hinting
the Iroshnikov-Kraichnan spectrum. These characteristics of
the VWF, WF, and SF classes are consistent with their incompressible
turbulence counterparts, indicating that a modest compressibility of
${\langle \delta\rho / \rho \rangle}_{\rm rms} \la 0.45$, or so,
does not play a significant role in turbulence. Our simulations were
able to produce the SF class behaviors only with high resolution
of at least $1024^2$ grid cells. With lower resolutions, we
observed the formation of a dominant flux tube, which accompanies
the separation of magnetic field from the background flow.
The specific requirements for the simulation of the SF class
should depend on the code (and the numerical scheme) as well as
the initial setup, but our results do indicate that very high
resolution would be required for converged results in
simulation studies of MHD turbulence.

\end{abstract}

\section {Introduction}

The existence of cosmic magnetic fields in diffuse astrophysical plasmas
(the interstellar media in galaxies, the intracluster media in clusters of
galaxies and even the media associated with filaments and superclusters
of galaxies) have been recognized for a while (for reviews, see \eg Kronberg
1994; Beck \etal 1996; Zweibel \& Heiles 1997; Carilli \& Taylor 2002;
also see Ryu \etal 1998; Clarke \etal 2001). Although the origin of such
fields is not yet fully understood, turbulence is known to play an important
role in the amplification and diffusion of existing magnetic fields.
For instance, magnetic fields grow effectively by turbulent motion
of conducting fluids (the $\alpha$ effect), but they are not further
amplified when Maxwell stresses become 
strong enough to affect the turbulent motion itself. How strong the
magnetic fields would be in order to influence turbulence is one of
the intriguing topics in magnetohydrodynamic (MHD) turbulence.
It has been argued through two-dimensional incompressible simulations that
turbulent transport is reduced by weak magnetic fields whose energy is
small compared to the kinetic energy of turbulent flows (Cattaneo \&
Vainshtein 1991; Cattaneo 1994). A similar suppression was also observed
in three-dimensional simulations of incompressible flows (Tao \etal 1993).
It is an important issue in astrophysics, since it is a part of
the process for the generation and evolution of cosmic magnetic fields
(see \eg Ruzmaikin \etal 1988; Kulsrud \etal 1997; Kulsrud 1999).

In this work, we study compressible MHD turbulence by solving
the ideal MHD equations with an isothermal equation of state.
Since turbulent motion produces structures spanning a wide range of
scales with accompanying energy transfer among different scales,
high spatial resolution is required to cover a sufficient inertial
range. Hence, in this paper we present high resolution two-dimensional 
simulations using up to $1536^2$ grid cells, leaving three-dimensional
simulations for follow-up work. In addition, in order to achieve
the highest possible magnetic Reynolds number and Reynolds number,
no dissipation was included explicitly in our simulations.
However, the resistivity and viscosity of numerical origin are still
effective  as small-scale dissipative channels. 
From the simulations, we examine the basic properties of
the resulting MHD turbulence. We also address the dependence of
the properties on resolution, including the amplification of
magnetic field.

The rest of the paper is organized as follows. In the next section,
we describe the problem setup as well as the code properties.
In \S 3 and \S 4, we present the results of simulations of hydrodynamic
and MHD turbulence. Finally in \S 5, the findings of this study are
summarized.

\section {Numerics}

In simulations of compressible MHD turbulence, gas is heated by shocks
and reconnection events, as well as numerical dissipation that mimics
viscous and resistive influences (see below). Hence, in order to
maintain the turbulence statistically in a steady-state, cooling should
be applied to the internal energy. One simple way to handle it is to assume
the isothermality of flows. The MHD equations of compressible,
isothermal gas are
\begin{equation}
\frac{\partial \rho}{\partial t} + {\vec \nabla}\cdot ({\rho \vec v}) = 0,
\label{continuity equation}
\end{equation}
\begin{equation}
\frac{\partial \vec v}{\partial t} + {\vec v} \cdot {\vec \nabla \vec v}
+ {c_s^2 \over {\rho}}{\vec \nabla}{\rho}  -
{1 \over {\rho}}{({\vec \nabla}\times{\vec B})\times{\vec B}} = {\vec f},
\label{momentum equation}
\end{equation}
\begin{equation}
\frac{\partial \vec B}{\partial t} - {\vec \nabla}
\times ({\vec v} \times {\vec B}) = 0,
\label{equation of magnetic field evolution}
\end{equation}
with an additional constraint 
\begin{equation}\label{divb}
\vec{\nabla} \cdot \vec{B}=0,
\end{equation}
for the absence of magnetic monopoles. Here, $c_s$ is the isothermal sound
speed, and the units are chosen so that the factor of $4\pi$ does not
appear in the equations. To enforce turbulence, a random force per
mass, $\vec f = f_x {\hat x} + f_y {\hat y}$, was added.
It has the following form:
\begin{equation}
{f_{x,y}(x,y,t)} = {\it v_{\rm amp}} \cos({\it \omega}t + {\delta_t})
\cos(\it k_{in, x} \it x+ {\delta_x}) \cos(\it k_{in,y} \it y + {\delta_y}),
\label{the random forcing function}
\end{equation}
where $\delta_t$, $\delta_x$, and $\delta_y$ are random phases in
the range of $0 \le {\delta_t}, {\delta_x}, {\delta_y} \le \pi$.

The above equations were solved using a multi-dimensional MHD code
described in Kim \etal (1999), which is specifically designed for
isothermal MHD. It is based on the explicit, finite-difference Total
Variation Diminishing (TVD) scheme, which is a second-order accurate
upwind scheme, and employs the minmod flux limiter. Simulations were
performed in the computational domain of $x = [0,L]$, $y = [0,L]$,
and $L=1$ with a periodic boundary condition using $n_x^2
= {256^2}$, ${512^2}$, ${1024^2}$ and ${1536^2}$ grid cells.
The total mass in the computational box is conserved, and the averaged
density was set to be $\langle\rho\rangle = \rho_0 = 1$.
The values of other parameters in the simulations are the following:
the isothermal sound speed $c_s=1$, the angular frequency of the random
forcing $\omega=2\pi$, and the input wavenumbers
$\it k_{in,x,y}$ = $8 \pi$. With this choice, we note that the input
scale of the random forcing is one quarter of the computational box
size and the period is one sound wave crossing time across the box.
The amplitude of the random forcing, $v_{amp}$, was set so that
without a magnetic field, hydrodynamic turbulence is saturated with
the averaged Mach number $\langle M_s \rangle_{\rm init} \equiv
\langle v \rangle_{\rm rms} / c_s = 1$ (see \S 3). MHD turbulence
was generated by introducing uniform magnetic fields of various
strength into the saturated hydrodynamic turbulence (see \S 4).

Although our simulations do not contain explicit resistivity or viscosity,
unavoidable numerical diffusion of magnetic field and momentum across cell
produce effective numerical resistivity and
viscosity, respectively. Kim \etal (1999) studied the character of
numerical dissipation in the code used for this work through the decay
of a two-dimensional Alfv\'en wave. They showed that for waves
spanning different numbers of grid cells, $n$, the effective magnetic
Reynolds number and Reynolds number are proportional to
$n^{1.66}$, mimicking the ``hyper-type'' resistivity and
viscosity ($\eta\nabla^4{\vec B}$ and $\mu\nabla^4{\vec v}$)
rather than the ``normal-type'' resistivity and viscosity
($\eta\nabla^2{\vec B}$ and $\mu\nabla^2{\vec v}$) that characterize
the physical dissipation of ``collisional'' fluids. They also
estimated that the effective Reynolds numbers are larger than several
hundreds if $n\ge8$, so that the inertial range covers scales spanning
8 grid cells or more. We note that numerical dissipation estimates
depend somewhat on the tests used. For instance, the test suggested
by Zweibel \etal (2002) would have given somewhat smaller numerical
dissipation. Hence, our estimation of the inertial range based on
the Alfv\'en wave decay can be regarded to be rather conservative.

In most numerical studies of astrophysical turbulence based on explicit
dissipation, the normal-type resistivity and viscosity have been 
included explicitly. However, dissipation processes in diffuse
astrophysical plasmas are not well understood, and it is unlikely
that the dissipation can be modeled by way of simple resistivity and
viscosity coefficients. Hence, more careful consideration would be
required in the studies of the properties of astrophysical turbulence
when details on dissipative scales become important. In this paper
we are concerned mostly with the properties of turbulence in
the inertial range, which would be less sensitive to the form of
dissipation, so we ignore dissipative complications outside this range.

\section{Hydrodynamic Turbulence}

Simulations of hydrodynamic turbulence were first performed for
two reasons. One is to generate the initial states for MHD simulations.
The other is to observe the change of turbulence properties from the the
hydrodynamic case under the influence of magnetic fields. In the absence
of magnetic fields, a trans-sonic turbulence with
$\langle M_s \rangle_{\rm init} \simeq 1$ developed (see Table 2).

Figure 1 shows a typical resulting density distribution at an epoch
$t=15$ and the time-averaged power spectrum of kinetic energy in
the simulation with $1536^2$ grid cells. The existence of weak shocks
is evident in the density image. Here, ``weak'' means the shock Mach
number is small; typically $M_{\rm shock} \sim 2$. The density
compression is $\langle\delta\rho/\rho\rangle_{\rm rms} \simeq 0.45$,
when averaged over space and time in all the simulations of all
resolutions (see Table 3). In a two-dimensional system, the forced
turbulence of ``incompressible'' flows exhibits the dual energy
spectrum known as the Kraichnan spectrum: in the range of $k < k_{in}$,
the slope follows $k^{-5/3}$ indicating the direct cascade of energy,
while in the range of $k > k_{in}$, it follows $k^{-3}$ meaning
the direct cascade of enstrophy (see \eg Lesieur 1997 and references
therein). In the case of ``high compressibility,'' where strong shocks
are common, the energy spectrum follows the $k^{-2}$ slope
in the inertial range, known as the Burgers spectrum (see \eg Lesieur
1997 and references therein). In our results with weak compressibility,
the energy spectrum is consistent with the Kraichnan spectrum as shown in
the right panel of Figure 1, but the slope for $k > k_{in}$ is $\sim 2.72$;
somewhat shallower than for incompressible turbulence. As listed in Table 4,
the deviation of the slope from $k^{-3}$ increases systematically as
the resolution increases. This is because in higher resolution simulations,
shocks are better resolved, and, hence, their effect is more evident
in the energy power spectrum.

\section{MHD Turbulence}

Weak, uniform magnetic fields of various strength were added to the fully
developed hydrodynamic turbulence. In the $n_x^2 = 256^2$,
$512^2$ and $1024^2$ simulations, the flows of hydrodynamic turbulence
at $t=30$ were taken as the initial states, while in the $1536^3$
simulations, the flow at $t=15$ (the right panel of Figure 1) was taken.
The strength of the added magnetic fields was set so that
the initial Alfv\'enic Mach number, $\langle M_A \rangle_{\rm init}
\equiv \langle v \rangle_{\rm rms} / c_{A,{\rm init}}$,
ranged from $1000$ to $10$ (it corresponds to the plasma $\beta$
of $2\times10^6$ to $200$). The Alfv\'enic Mach number
represents the ratio of Reynolds stresses to Maxwell stresses, so is
a convenient way to characterize magnetic field strength in these flows.

The magnetic field strength grew rapidly, almost exponentially,
during the initial transient period, and then saturated.
Figure 2 shows the time evolution of rms magnetic field strength and flow
velocity. The resulting flows of MHD turbulence after saturation
can be naturally categorized into {\it very weak field} (VWF),
{\it weak field} (WF) and {\it strong field} (SF) classes (Table 1).
In the VWF cases, magnetic field influences are negligible, and
the flow velocity does not change noticeably from purely hydrodynamic
turbulence. In the WF cases, the magnetic field energy, $E_{mag}$,
is still small compared to the flow kinetic energy, $E_{kin}$, but
the magnetic field does locally affect small scale flow motions. Hence,
the turbulence flow velocity is decreased from that of non-magnetic cases.
Finally, the cases where $E_{mag}$ becomes comparable to $E_{kin}$ are
classified as the SF class. We note that the change in the flow properties
is gradual with the increase of the strength of the initial, uniform fields.
Hence, boundaries between the classes are somewhat arbitrary.
We set the criterion for the boundary between the VWF
and WF classes as $\langle M_s \rangle \equiv \langle v
\rangle_{\rm rms} / c_s \la 1$; more specifically, $\langle M_s \rangle
= 0.9$ was taken. On the other hand, the criterion for the boundary
between the WF and SF classes was set as
$\langle M_A \rangle \equiv \langle v \rangle_{\rm rms} /
\langle c_A \rangle_{\rm rms} \ga 1$, more specifically
$\langle M_A \rangle =2$ was taken. Among the SF cases,
we observed examples where there a change in the magnetic field
configuration takes place. In particular there are cases in which
a dominant flux tube develops. These cases we categorized specifically 
as the {\it field separation} (FS) class. The growth of magnetic field
strength and the characteristics of the four classes are further detailed
in the following subsections.

For quantitative discussions we computed, in addition to
$\langle M_A \rangle$ and $\langle M_s \rangle$ (Table 2), the density
compression, ${\langle \delta\rho / \rho \rangle}_{\rm rms}$, and
the intermittency (or the kurtosis of the field distribution)
$I \equiv \langle B^4 \rangle_{\rm rms} / \langle B^2 \rangle_{\rm rms}^2$
(Table 3), as well as the slopes of the power spectra of kinetic and
total energies (Table 4). Note that with a Gaussian distribution of
magnetic field, ${\vec B} \propto \exp(-x^2/2\sigma^2) {\hat y}$,
the value of $I$ is 2.39 when summed over the interval
$-3\sigma \le x \le 3\sigma$, or 3.99 when $-5\sigma \le x \le 5\sigma$
is taken into account.

\subsection{Growth of Magnetic Field Strength}

There is no dynamo action in two-dimensions, and the total magnetic flux
through given boundaries is conserved. Hence, the growth of magnetic field
strength shown in Figure 2 is due to stretching and compression. The
growth stops either when the magnetic energy reaches an equipartition
with the kinetic energy, so that the back-reaction from magnetic field
Maxwell stresses plays a dominant role, or when the separation between
magnetic sheets is reduced to the diffusive scale, so that reconnection
takes place (see \eg Biskamp 1993 for details). If the initial,
uniform magnetic field is weak enough, the growth is saturated by
the latter cause before the magnetic energy reaches equipartition, as
in the VWF and WF cases. Then, the resulting growth should depend on
the diffusive scale and, hence, on the effective resistivity. Then the
amplification of magnetic energy from the initial value is expected to
be $\la {\it (k_{in} \delta)}^{-2} \sim {R_m}$ (see \eg Biskamp 1993).
Here, $\delta$ is the diffusive scale and $R_m$ is the magnetic
Reynolds number.

Figure 3 shows the averaged amplification of magnetic energy in our
simulations as a function of grid resolution. Only the VWF and WF
cases were considered, because in the SF cases, the growth of magnetic
field stopped as a consequence of Maxwell stresses, rather than
reconnection. The amplification shown in the figure is proportional
to $n_x^{1.32}$. Furthermore, with $R_m \propto n^{1.66}$ for
our code (see \S 2), the amplification is estimated to be approximately
proportional to ${R_m}^{0.8}$, which is somewhat shallower than the expected
dependence ($\propto {R_m}$). However, this result agrees well with results
of two-dimensional numerical simulations for incompressible MHD turbulence
(Biskamp 1993).

\subsection{VWF Cases}

The VWF class includes those cases where the back-reaction of magnetic
field into flow motions is insignificant. The criterion
$\langle M_s\rangle \ga 0.9$ corresponds to the rms velocity decrease
by less than 10\%. The characteristics of turbulence
in these cases are summarized as follows: 1) The density
compression is still large with ${\langle\delta\rho
/\rho\rangle}_{\rm rms} \ga 0.38$ (Table 3) and shocks still exist
as shown clearly in the bottom-left panel of Figure 4. 2). The
intermittency is large with $I \ga 3.5$ (Table 3), which indicates that
the spatial contrast in the magnetic field strength distribution
is high. In particular, the bottom-right panel of Figure 4 shows that
the magnetic field is mostly thin tubes (or thin sheets in extension to
the third dimension). 3) The Alfv\'en Mach number is $\langle
M_A\rangle \la 4.5$ (Table 2), and hence the kinetic energy is,
at least, an order of magnitude larger than the magnetic energy.
Also, the power of kinetic energy, ${P_k}^{kin}$, is larger than
that of magnetic energy, ${P_k}^{mag}$, over most wavenumbers,
as shown in the bottom panels of Figure 5. However, ${P_k}^{mag} \ga
{P_k}^{kin}$ in a small range of large wavenumbers. This is because
the magnetic field first built up on small scales. However, it saturated
due to reconnection before the power could extend to larger scales.
4) The slope of ${P_k}^{kin}$ was not affected much by the magnetic
field. The change was less than a few percent as shown in Table 4.
But the slope of the power of the total energy, ${P_k}^{tot}$, changed
significantly compared to the hydrodynamic case, especially in the MHD
cases close to the boundary between the VWF and WF classes.
This is because the small scale power of magnetic energy pushed
${P_k}^{tot}$ up, and hence ${P_k}^{tot}$ does not have a single,
well-defined slope any more.

\subsection{WF Cases}

The WF class was categorized by the property that globally the
magnetic energy is smaller than the kinetic energy,
but yet the back-reaction of magnetic field is not negligible. As the
criterion to represent this property, $\langle M_s\rangle \la 0.9$ and
$\langle M_A\rangle \ga 2$ were adopted (Table 2).
The characteristics of turbulence in these cases are the following:
1) As shown in the middle-left panel of Figure 4, the occurrence
of shocks is reduced and their strength is weakened compared to
the hydrodynamic case. The density compression is
$0.28 \la {\langle \delta\rho / \rho\rangle}_{\rm rms} \la 0.38$
(Table 3), which is smaller than that in the VWF cases. 2) The contours
of magnetic field lines plotted in the middle-right panel of Figure 4
show the appearance of almost circular magnetic islands. At the same time, the
magnetic field is less concentrated than in the VWF cases, with smaller
intermittency $2.2 \la I \la 3.5$ (Table 3).
3) Although in the small wavenumber region the magnetic energy power 
is still smaller than the kinetic energy power, $P^{mag}_k \ll {P_k}^{kin}$, 
at large wavenumbers, ${P_k}^{mag} \ga {P_k}^{kin}$, as shown
in the middle panels of Figure 5. Hence, turbulent transport can
be suppressed by magnetic fields on the small scales where ${P_k}^{mag}
\ga {P_k}^{kin}$. Consequently, the flow velocity decreases noticeably from
that of hydrodynamic turbulence (see Figure 2). 4) The power of
magnetic energy, ${P_k}^{mag}$, changed the shape of
${P_k}^{tot}$ as well as that of ${P_k}^{kin}$ in power spectra,
and the slopes are not simply defined over the entire inertial
range of wavenumber any more (see Figure 5). We calculated the
slopes of ${P_k}^{kin}$ and ${P_k}^{tot}$ over $2\times k_{in} \le k
\le k_{24}$, where the slopes are relatively well defined, and
listed them in Table 4. Here, $k_{24}$ is the wavenumber corresponding
to 24 grid cells. The slope of ${P_k}^{kin}$ is significantly smaller
than those of the hydrodynamic and VWF cases. The slope of
${P_k}^{tot}$ is quite small. But this is because ${P_k}^{mag}$ peaks
in the wavenumber range over which the slope was calculated (see Figure 5).

\subsection{SF Cases}

In the SF cases, the magnetic energy grows to become comparable to the
kinetic energy, with $\langle M_A\rangle \la 2$ at saturation (Table 2).
Hence, the flow velocity is influenced significantly by magnetic field
(Figure 2), with $\langle M_s\rangle \la 0.75$ at saturation (Table 2).
The characteristics of SF turbulence are the following: 1) With
the significantly reduced Mach number, shocks are rare, as shown in
the top-left panel of Figure 4. The density compression is accordingly
reduced to ${\langle \delta\rho / \rho\rangle}_{\rm rms} \la 0.28$
(Table 3). 2) As shown in the top-right panel of Figure 4,
circular magnetic flux islands, or loops, are common. This trend of changing
topology in magnetic field lines, from tubes to loops, with increase of
magnetic field strength was observed also in the two-dimensional
simulations of incompressible MHD turbulence (see \eg Biskamp 1993).
At the same time, the spatial contrast of magnetic field strength
distribution becomes low with smaller intermittency $I \la 2.2$ (Table 3).
3) The power of magnetic energy exceeds that of kinetic energy
with $P_k^{mag} \ga P_k^{kin}$ over all $k \ga  k_{in}$ as shown
in the top panels of Figure 5, although still $P_k^{mag} < P_k^{kin}$ in
$k \la k_{in}$. 4) The slopes of ${P_k}^{kin}$ and ${P_k}^{tot}$,
calculated over $2\times k_{in} \le k \le k_{24}$ as in the WF cases,
are listed in Table 4. Across the above wavenumber range, ${P_k}^{kin}$,
${P_k}^{tot}$ and ${P_k}^{mag}$ all exhibit a single slope close to
$\sim 1.5$; namely, the slope of the Iroshnikov-Kraichnan spectrum
(the top panels of Figure 5).

We note that the emergence of the slope of $\sim 1.5$ is
achieved only in the very high resolution simulations with $1024^2$ and
$1536^2$ grid cells, not in the simulations with $256^2$ and $512^2$ grid
cells. In fact, there is no case categorized as the SF class in the
simulations with $256^2$ and $512^2$ grid cells at all (see Table 1).
That is, our two-dimensional simulations of MHD turbulence start to
show converged behaviors with $1024^2$ grid cells or more.

\subsection{FS Cases}

The cases with $\langle M_A \rangle_{\rm init} \la 30$ show FS class
behavior, wherein a dominant flux tube forms, and, as a consequence,
the magnetic field separates from the background flow (see Fig. 6).
In such FS cases, flows become anisotropic and their characteristics
are different from the other cases. The appearance of this behavior
can be understood as follows. Recall that there are three ideal 2D
MHD invariants; the total energy, $E_{tot}$, the mean square of
the magnetic potential, $A$, and the cross helicity,
\begin{equation}
K = {1 \over 2} \int {\vec v}\cdot{\vec B} d^2{\vec x}.
\end{equation}
Among them, the power spectrum of $A$ exhibits an inverse cascade,
while the power spectra of $E_{tot}$ and $K$ exhibit a normal cascade
(see \eg Biskamp 1993). Hence, just as in three dimensions, large scale
magnetic field power can be built up in the regime $k \le k_{in}$ in
two-dimensional MHD turbulence. In our FS cases this power build-up
proceeds through the formation of a magnetic flux tube with a coherent
length larger than the scale associated with $k_{in}$. 
However, as pointed out in \S 4.1, there is no dynamo action
in our simulations. Hence, the formation of such a flux tube should involve
reconnection followed by the expulsion of gas out of the flux tube,
reducing the gas density (so, also the gas pressure) inside the tube,
while maintaining approximate pressure equilibrium.

We note that the critical value of $\langle M_A \rangle_{\rm init}$ for
the FS class is independent of numerical resolution in our simulations
(see Table 1). This is because the existence of the flux tube depends
on its ability to resist ram pressure bending, which depends on large
scale flows, so is independent of numerical resolution. On the other hand,
since the formation of the dominant flux tube involves reconnection,
we expect the time to reach the state where the magnetic field separates
from the background flow would depend on resistivity. The flux tube was
developed in a few tens eddy turnover times in our simulations. However,
in astrophysical environments such as interstellar media or intracluster
media, the resistivity is much smaller than that of our simulations. The
classical magnetic Reynolds number in the interstellar medium can be as
large as $10^{50}$ or so (see \eg Spitzer 1979). Hence, we expect that the
emergence of the flux tube would take too long in astrophysical environments,
and instead, turbulence of the SF class is more likely to develop. So the
failure of SF cases in our $256^2$ and $512^2$ simulations is probably
a limitation of numerical simulations; in particular from very large
numerical resistivity compared to astrophysical environments and perhaps
also from the periodic boundary we employed.

\section {Summary \& Conclusion}

We performed high resolution two-dimensional simulations of isothermal
MHD turbulence using up to $1536^2$ grid cells. Compressibility was taken
into account, with Mach numbers $\langle M_s \rangle \la 1$. In order to
maximize the magnetic Reynolds number and the Reynolds number,
no explicit dissipation was included. Instead, resistivity and viscosity
of numerical origin were utilized. Our findings are summarized as follows:\\
1) The growth of magnetic energy from the initial value depends on
effective resistivity and, hence, on  resolution. We measured a maximum
magnetic energy amplification proportional to $n_x^{1.32}$. 
After the growth saturated, the turbulence could be categorized into
three classes, which we labeled very weak field (VWF), weak field (WF),
and strong field (SF) classes, depending on the strength of the uniform
component of magnetic fields, or equivalently the initial Alfv\'en Mach
number, $\langle M_A \rangle_{\rm init}$. Each class is characterized
by different turbulence properties.\\
2) In the WF cases, although $E_{mag} < E_{kin}$ globally, the rms
flow velocity decreases due to the magnetic field, which can become
locally important. That is, even a weak magnetic field reduces
turbulent transport (Cattaneo \& Vainshtein 1991; Cattaneo 1994).\\
3) Although the inertial range power spectra of MHD turbulence are not
represented by power-laws any more, part of the inertial range still
develops a slope close to $\sim 1.5$; that of the Iroshnikov-Kraichnan
spectrum, in the SF cases. Such a slope developed only in highest
resolution simulations with $1024^2$ and $1536^2$ grid cells.\\
4) In the lower resolution simulations with $256^2$ and $512^2$ grid cells,
there are no cases categorized as the SF class. Instead, with
$\langle M_A \rangle_{\rm init} \la 30$, those simulations developed into
FS cases, where a large flux tube dominates the flow structures and
separates the flow from the magnetic field. This FS property is the 
result of large numerical resistivity along with the periodic box used.

The conclusions of our work are the following:\\
1) For the VWF, WF, and SF cases, we observed properties of turbulence
which are consistent with those of incompressible counterparts, whenever
comparisons are made. Hence, we conclude that weak compressibility of
${\langle \delta\rho / \rho \rangle}_{\rm rms}$ up to $\sim0.45$
would not be important in characterizing MHD turbulence.\\
2) Converged behavior in simulated turbulence, such as that of the SF class,
starts to appear only in simulations with very high resolution, $1024^2$
grid cells or more, in our simulations. The fact that simulations with
$1024^2$ grid cells or more are necessary for our two-dimensional study
of MHD turbulence points to the need of very high resolution in
three-dimensional MHD turbulence studies as well. For hydrodynamic
turbulence, Porter \etal reported that converged results emerged with
$512^3$ or more grid cells in their simulations. However, we note that
numerical dissipation would differ in different codes based on different
schemes. As a result, the convergence behavior would be different, too.
These findings emphasize the importance in turbulence studies of high
resolution grids and codes with the smallest possible numerical
dissipation. Any explicit treatment of dissipation will require even
higher resolution simulations, since numerical dissipation and diffusion
would have to be smaller than the explicit terms.

\acknowledgments

The work by HL and DR was supported by an international collaboration
grant from Korea (KOSEF F01-2000-000-10009-0) and Germany (DFG).
The work by JK was supported by Strategic National R\&D Program
(M10222000005-02B0600-00400) from MOST, Korea.
The work by TWJ was supported by the National Science Foundation
(AST96-19438; AST00-71167) and the University of Minnesota
Supercomputing Institute. Simulations were made through the support
by ``The Supercomputing Application Support Program of KISTI''.

\clearpage

\begin{deluxetable}{ccccccccc}
\tabletypesize{\footnotesize}
\tablenum{1}
\tablecaption{Classification of turbulence/Time at the end of simulations,
$t_{\rm end}$}
\tablehead{ \colhead{$\langle M_A \rangle_{\rm init}$} & \colhead{10} & \colhead{20} & \colhead{50} & \colhead{100} & \colhead{200} & \colhead{300} & \colhead{1000}& \colhead{$\infty$}}
\startdata
${512}^2$~\tablenotemark{a}&FS\tablenotemark{b}~/80&FS\tablenotemark{b}~/100&\underline{WF}\tablenotemark{b}~/40&VWF\tablenotemark{b}~/40&VWF\tablenotemark{b}~/40&VWF\tablenotemark{b}~/40&VWF\tablenotemark{b}~/40&HD\tablenotemark{b}~/40\\
\\
${1024}^2$~\tablenotemark{a}&FS\tablenotemark{b}~/110&FS\tablenotemark{b}~/140&SF\tablenotemark{b}~/40&\underline{WF}\tablenotemark{b}~/40&VWF\tablenotemark{b}~/40&VWF\tablenotemark{b}~/40&VWF\tablenotemark{b}~/40&HD\tablenotemark{b}~/40\\
\\
${1536}^2$~\tablenotemark{a}& & &SF\tablenotemark{b}~/20&\underline{WF}\tablenotemark{b}~/20&\underline{WF}\tablenotemark{b}~/15&VWF\tablenotemark{b}~/20& &HD\tablenotemark{b}~/20\\
\enddata
\tablenotetext{a}{The number of grid cells used. The simulations with
$256^2$ grid cells were performed, but are not listed in tables.}
\tablenotetext{b}{Standing for field separation, strong field, weak field,
very weak field, and hydrodynamic. See the text for details.}
\end{deluxetable}

\begin{deluxetable}{ccccccc}
\tablenum{2}
\tablecaption{Alfv\'en Mach number $\langle M_A \rangle$/Mach number
$\langle M_s \rangle$ for SF, WF, and VWF cases~\tablenotemark{a}}
\tablehead{ \colhead{$\langle M_A \rangle_{\rm init}$} & \colhead{50} & \colhead{100} & \colhead{200} & \colhead{300} & \colhead{1000} & \colhead{$\infty$\tablenotemark{b}}}
\startdata
${512}^2$&\underline{2.38/0.83}&4.93/0.95&9.70/0.99&14.90/1.00&49.66/1.00&$\infty$/1.00\\
\\
${1024}^2$&1.60/0.70&\underline{3.02/0.86}&6.00/0.95&9.05/0.98&31.38/1.00&$\infty$/1.00\\
\\
${1536}^2$&1.41/0.66&\underline{2.28/0.81}&\underline{4.04/0.85}&6.57/0.97& &$\infty$/1.00\\
\enddata
\tablenotetext{a}{Averaged spatially over the whole computational domain,
and temporally over $20\le t\le t_{\rm end}$ for $512^2$ and $1024^2$
simulations and over $10\le t\le t_{\rm end}$ for ${1536}^2$ simulations.
The values for the WF cases are underlined for clarity.}
\tablenotetext{b}{Hydrodynamic turbulence.}
\end{deluxetable}

\begin{deluxetable}{ccccccc}
\tablenum{3}
\tablecaption{Density compression $\langle \delta\rho/\rho
\rangle_{\rm rms}$/intermittency $I$ for SF, WF, and VWF
cases~\tablenotemark{a}}
\tablehead{ \colhead{$\langle M_A \rangle_{\rm init}$} & \colhead{50} & \colhead{100} & \colhead{200} & \colhead{300} & \colhead{1000} & \colhead{$\infty$\tablenotemark{b}}}
\startdata
${512}^2$&\underline{0.32/2.68}&0.39/3.86&0.42/4.80&0.42/5.28&0.45/5.93&0.45/***\\
\\
${1024}^2$&0.25/2.06&\underline{0.34/3.03}&0.40/4.28&0.42/5.05&0.44/5.88&0.45/***\\
\\
${1536}^2$&0.24/1.95&\underline{0.30/2.48}&\underline{0.37/3.48}&0.41/4.58& &0.45/***\\
\enddata
\tablenotetext{a}{Averaged spatially over the whole computational domain,
and temporally over $20\le t\le t_{\rm end}$ for $512^2$ and $1024^2$
simulations and over $10\le t\le t_{\rm end}$ for ${1536}^2$ simulations.
The values for the WF cases are underlined for clarity.}
\tablenotetext{b}{Hydrodynamic turbulence.}
\end{deluxetable}

\begin{deluxetable}{ccccccc}
\tablenum{4}
\tablecaption{Spectral slope of kinetic energy/total energy
for SF, WF, and VWF cases~\tablenotemark{a}}
\tablehead{ \colhead{$\langle M_A \rangle_{\rm init}$} & \colhead{50} & \colhead{100} & \colhead{200} & \colhead{300} & \colhead{1000} & \colhead{$\infty$\tablenotemark{b}}}
\startdata
${512}^2$&\underline{2.22/1.27}&2.77/2.21&2.95/2.78&2.67/2.61&2.77/2.76&2.90/2.90\\
\\
${1024}^2$&1.46/1.27&\underline{2.07/1.20}&2.74/2.09&2.84/2.51&2.75/2.71&2.85/2.85\\
\\
${1536}^2$&1.45/1.48&\underline{1.68/1.18}&\underline{2.33/1.50}&2.68/2.05& &2.72/2.72\\
\enddata
\tablenotetext{a}{Averaged temporally over $20\le t\le t_{\rm end}$ for
$512^2$ and $1024^2$ simulations and over $10\le t\le t_{\rm end}$ for
${1536}^2$ simulations. Slops was fitted over the range of
$2\times k_{in} \le k \le k_{24}$. Here, $k_{in}$ and $k_{24}$ are the
wavenumbers corresponding to the power input scale and 24 grid cells,
respectively. The values for the WF cases are underlined for clarity.}
\tablenotetext{b}{Hydrodynamic turbulence.}
\end{deluxetable}

\clearpage

\begin{figure}
\vspace{10.cm}
\figcaption{{\it Left panel} - Grey scale image of density for the
non-magnetic case with $1536^2$ resolution. Brighter regions correspond
to higher values and the gray scale was set arbitrarily to highlight
structures. The image represents the density distribution typical at
an epoch $(t=15)$ after hydrodynamic turbulence is fully developed.
{\it Right panel} - Power spectrum of kinetic energy, time-averaged
over $10\le t\le t_{\rm end}$, for the non-magnetic case with ${1536^2}$
resolution. Here, $k \equiv \sqrt{k_x^2 + k_y^2}$. The peak at
$k=5.7\times 2\pi$ corresponds to the wavenumber of the power input scale,
$k_{in} \equiv \sqrt{k_{in,x}^2 + k_{in,y}^2}$. For reference two vertical
lines are drawn at $k=2\times k_{in}$ and the wavenumber corresponding
to 8 grid cells, $k_8$. Two lines with slopes of $-5/3$ and $-3$ are
drawn for comparison.}
\end{figure}

\clearpage

\begin{figure}
\epsfxsize=16.5truecm
\centerline{\epsfbox{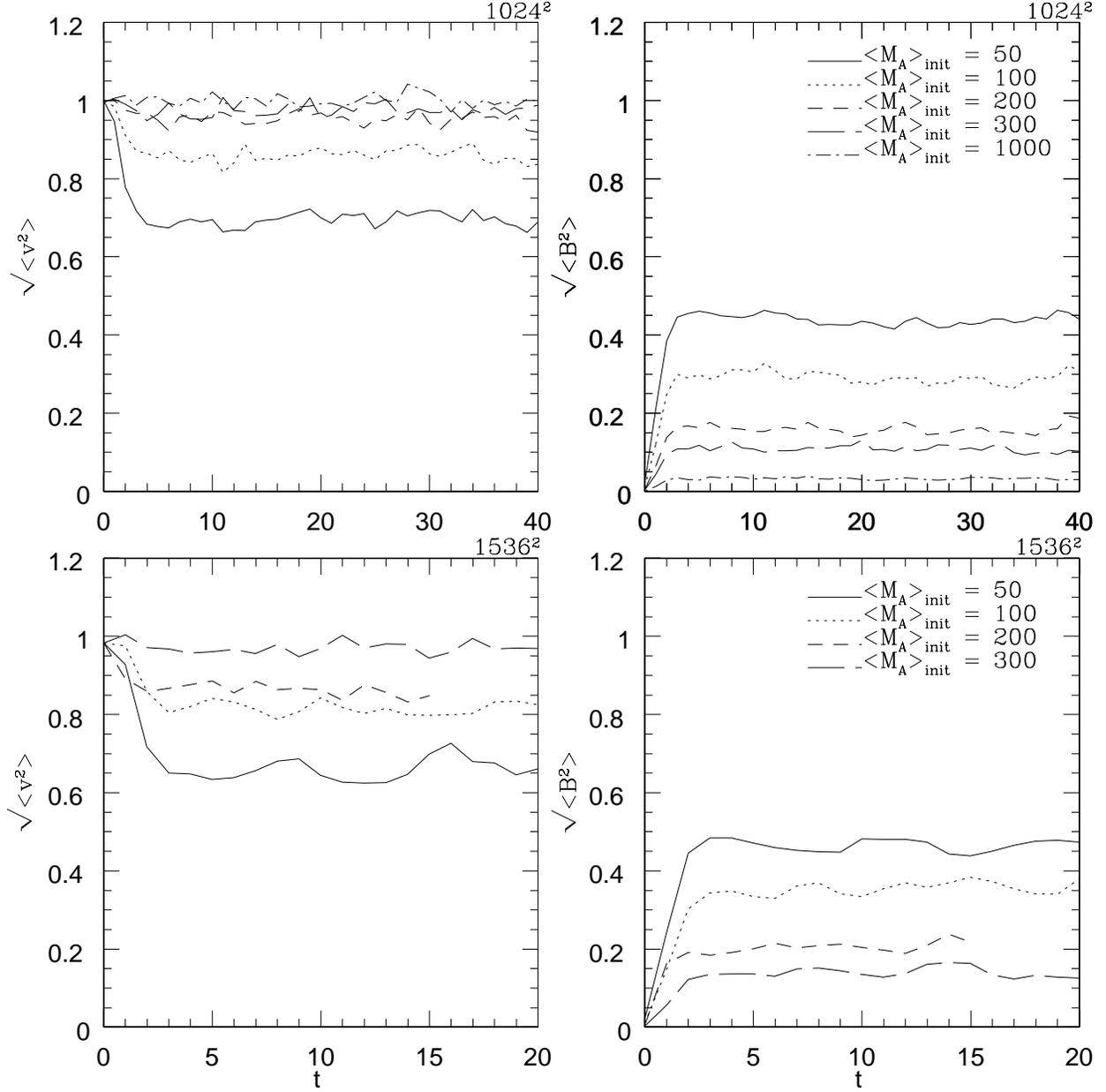}}
\figcaption{Time evolution of rms velocity (left panels) and magnetic
field strength (right panels), which were averaged over one period of
random forcing $(\tau=1)$, for different initial Alfv\'enic Mach numbers.
Upper panels are for the cases with $1024^2$ resolution and lower 
panels are for the cases with $1536^2$ resolution.}
\end{figure}

\clearpage

\begin{figure}
\epsfxsize=20truecm
\centerline{\epsfbox{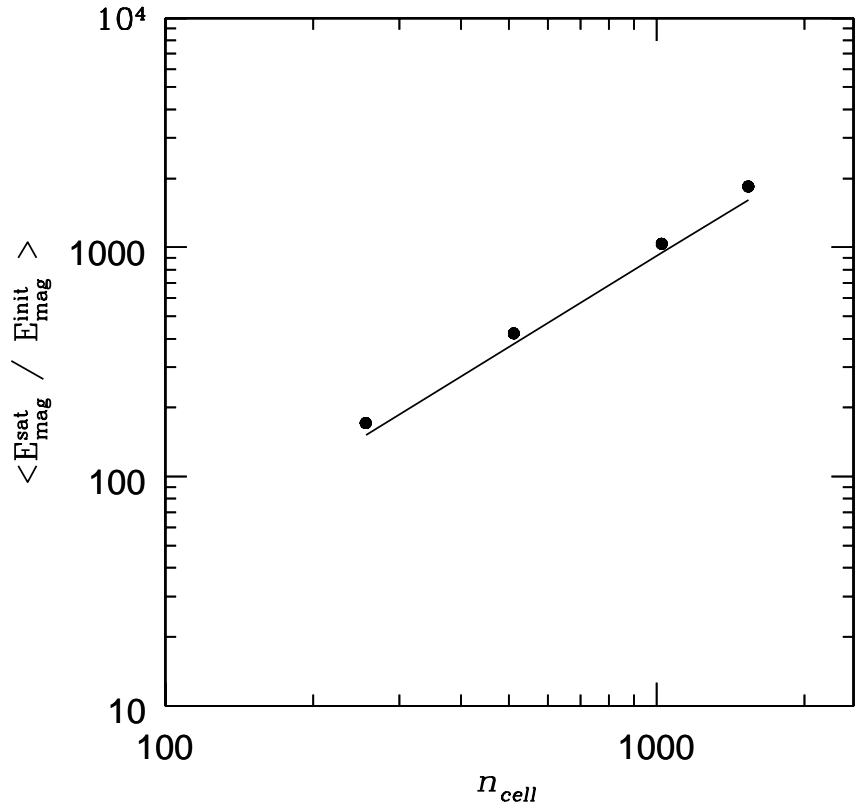}}
\vskip -6cm
\figcaption{Averaged amplification of magnetic energy from the initial
value in the VWF and WF cases as a function of numerical
resolution. Four points of $256^2$, $512^2$, $1024^2$, and $1536^2$ grid
cells are plotted. The fitted line $\propto n_x^{1.32}$ is drawn.}
\end{figure}

\clearpage

\begin{figure}
\vspace{10.cm}
\figcaption{Grey scale images of density (left panels) and contours of magnetic
field lines (right panels) at an epoch $(t = 13)$ for cases with $1536^2$
resolution. Upper two panels are for the case of
$\langle M_A \rangle_{\rm init} = 50$ classified as the SF class, middle
two panels are for $\langle M_A \rangle_{\rm init} = 100$ classified as
the WF class, and lower two panels are for
$\langle M_A \rangle_{\rm init} = 300$ classified as the VWF class.
In the density images, brighter regions represent higher values and
the gray scale was set arbitrarily to highlight structures.}
\end{figure}

\clearpage

\begin{figure}
\epsfxsize=18truecm
\centerline{\epsfbox{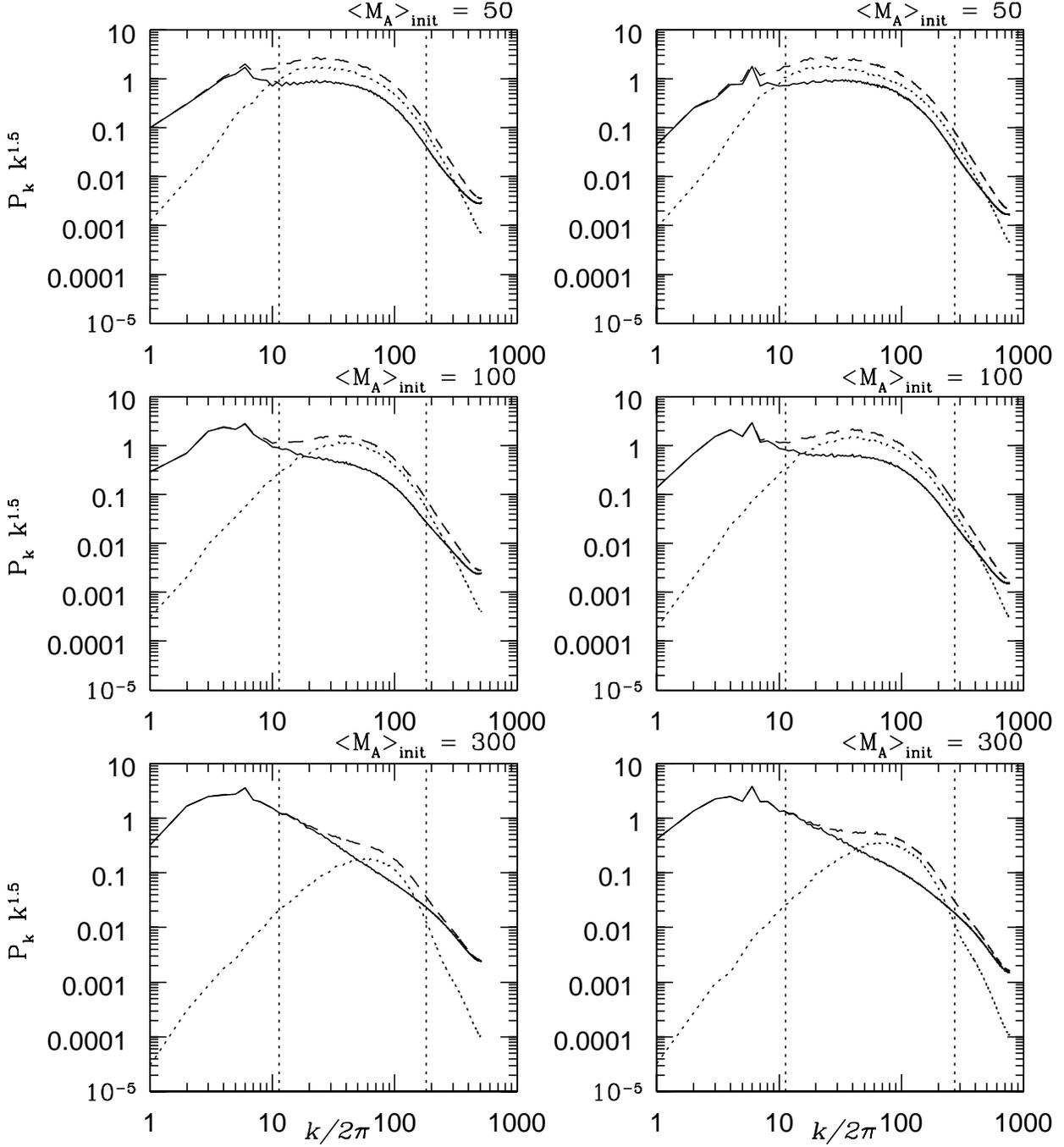}}
\figcaption{Power spectra, time-averaged over $20\le t\le t_{\rm end}$
for $1024^2$ resolution (left panels) and over $10\le t\le t_{\rm end}$
for $1536^2$ resolution (right panel). Spectra are multiplied by $k^{1.5}$
for clarity.
Solid line - power spectrum of kinetic energy, $P_k^{kin}$,
dotted line - power spectrum of magnetic energy, $P_k^{mag}$, and
dashed line - power spectrum of total energy, $P_k^{tot}$. From
top to bottom correspond to the SF, WF, and VWF classes,
respectively. The slopes fitted over the range of
$2\times k_{in} \le k \le k_{24}$ are shown in Table 4.
Here $k_{24}$ is the wavenumbers corresponding to 24 grid cells.}
\end{figure}

\clearpage

\begin{figure}
\vspace{10.cm}
\figcaption{Contours of magnetic field lines at four different times
for the case of $\langle M_A \rangle_{\rm init} = 10$ with $1024^2$
resolution. It belongs the FS class.}
\end{figure}


\begin{thebibliography}{}

\bibitem[Beck \etal(1996)]{bbmss96}
Beck, R., Brandenburg, A., Moss, D., Shukurov, A. \& Sokoloff, D.
1996, \araa, 34, 155
\bibitem[Biskamp(1993)]{bisk93}
Biskamp, D. 1993, Nonlinear Magnetohydrodynamics (Cambridge: Cambridge
Univ. Press)
\bibitem[Carilli \& Taylor(2002)]{ct02}
Carilli, C. L. \& Taylor, G. B. 2002, \araa, 40, in press (astro-ph/0110655)
\bibitem[Cattaneo(1994)]{catt94}
Cattaneo, F. 1994, \apj, 434, 200
\bibitem[Cattaneo \& Vainshtein(1991)]{cv91}
Cattaneo, F. \& Vainshtein, S. I. 1991, \apjl, 376, L21
\bibitem[Clarke \etal(2001)]{ckb01}
Clarke, T. E., Kronberg, P. P. \& B\"ohringer, H. 2001, \apjl, 547, L111
\bibitem[Kim \etal(1999)]{krjh99}
Kim, J., Ryu, D., Jones, T. W. \& Hong, S. 1999, \apj, 514, 506
\bibitem[Kronberg(1994)]{kron94}
Kronberg, P. P. 1994, Rep. Prog. Phys. 57, 325
\bibitem[Kulsrud (1999)]{kuls99}
Kulsrud, R. M. 1999, \araa, 37, 37
\bibitem[Kulsrud \etal(1997)]{kcor97}
Kulsrud, R. M., Cen, R., Ostriker, J. P. \& Ryu, D. 1997, \apj, 480, 481
\bibitem[Lesieur 1997]{lesi97}
Lesieur, M. 1997, Turbulence in Fluids, 3nd ed. (Dordrecht: Kluwer)
\bibitem[Porter \etal(1998)]{ppw98}
Porter, D., Pouquet, A. \& Woodward, P. 1998, Phys. Fluid, 10, 237
\bibitem[Ryu \etal(1998)]{rkb98}
Ryu, D., Kang, H., \& Biermann, P. L. 1998, \aap, 335, 19
\bibitem[Ruzmaikin \etal(1988)]{rss88}
Ruzmaikin, A.A., Shukurov, A.M. \& Sokoloff, D.D.
1988, in Astrophysics and Apace Science Library, Magnetic Fields in Galaxies,
(Dordrecht: Kluwer)
\bibitem[Spitzer(1979)]{spit79}
Spitzer, L. 1979, Physical Processes in the Interstellar Medium
(New York: Wiley-Interscience)
\bibitem[Tao \etal(1993)]{tcv93}
Tao, L., Cattaneo, F. \& Vainshtein, S. I. 1993, in Theory of Solar and
Planetary Dynamos, ed. M. R. E Proctor, P. C. Matthews \& A. M. Rucklidge
(Cambridge: Cambridge Univ. Press), p. 303
\bibitem[Zweibel \& Heiles(1997)]{zh97}
Zweibel, E. G. \& Heiles, C. 1997, Nature, 385, 131
\bibitem[Zweibel \etal(2002)]{zhf02}
Zweibel, E. G., Heitsch, F. \& Fan, Y. 2002, in Simulations of
Magnetohydrodynamic Turbulence in Astrophysics, ed. T. Passot \& E.
Falgarone (Berlin: Springer), in press (astro-ph/0202525)

\end{thebibliography}
\end{document}